\documentstyle[aps,preprint,epsf]{revtex}
\tightenlines

\newcommand\beq{\begin{equation}}
\newcommand\eeq{\end{equation}}

\begin{document}
\draft 
 
\title{
Kaon condensation at finite temperature}
\author{Toshitaka Tatsumi and Masatomi Yasuhira\\
Department of Physics, Kyoto University\\
Kyoto 606-01, Japan}

\maketitle
 
\begin{abstract}
A new formulation is presented to treat thermal fluctuations around 
the kaon condensate, based on chiral symmetry. Separating the zero
mode from the beginning we
perform the imaginary-time path integral to the one loop to get the
thermodynamic potential at finite temperature. The
role of the Goldstone mode in the kaon condensed phase
is elucidated in relation to the equation of state.


 
\end{abstract}
\newpage

Kaon condensation has been extensively studied for years
\cite{lee}. As their implications there have been suggested many
phenomena in relation to neutron stars. Among them the low-mass
black-hole scenario, proposed by Brown and Bethe \cite{bb},  
gives one of most interesting possibilities  for explaining the
current observation on the mass of neutron stars or the would-be
neutron star in the SN1987A. According to this scenario a newly formed
hot neutron star after the supernova explosion may collapse to a black
hole (the delayed collapse) during the deleptonization or the initial cooling 
era in the star's first
$\sim 20$s due to the softening of the equation of state
(EOS). A numerical simulation has been already
done following this scenario by Baumgarte et al.\cite{bau}. 
Through the studies of EOS it has been found that kaon
condensation results in the large softening of EOS and 
a low maximum mass for neutron stars. It may widely lie from 1.4$M_\odot$ to 
2.1$M_\odot$ in a relativistic calculation by Fujii et
al. \cite{tat}. A nonrelativistic calculation by Thorsson et al
also suggested similar values \cite{thor}. Anyway kaon condensation
gives a possibilty of a low maximum mass of $\sim 1.4M_\odot$, and
thereby the
low mass black holes can be expected around $1.4M_\odot$ \cite{bb}.

EOS for the kaon condensed phase has been studied mostly for low
temperature. However, the delayed collapse may occur in the hot 
protoneutron-star regime at temperature, $T \sim$ several tens of MeV, as
suggested in the simulations \cite{bau,kei}. Hence the studies of kaon
condensation at finite temperature is required.
Since there was no EOS of the kaon condensate at finite temperature, 
Baumgarte et al. used EOS at $T\simeq 0$ in their numerical
simulation \cite{bau}. 
Recently, Prakash et al. treated the
kaon-condensed phase at finite temperature and discussed the
properties of protoneutron stars within the meson exchange model,
since there is no consistent theory based on chiral symmetry
\cite{pra}.

Chiral symmetry is an important concept for kaon condensation; the
kaon condensed phase can be represented as a chirally-transformed
state from the vacuum in terms of chiral symmetry, and the essential
results can be obtained in a model-independent way \cite{tat}. So we
wish a consistent treatment of the thermal fluctuations based on
chiral symmetry even at finite temperature. In a recent paper 
Thorsson and Ellis have tried to include quantum or thermal
fluctuations within a chiral theory \cite{tho}. However, 
they have discussed only the
effect of the zero-point energy of kaons (quantum
fluctuations). Furthermore their expression for the thermodynamic
potential looks too complicated to be tractable.

In this Letter we present a new formalism to treat thermal
fluctuations on the basis of chiral symmetry and demonstrate how
our formalism works by calculating EOS. We shall see that our
formalism makes the analysis of the structure about the kaon dynamics
possible and clarifies the physics included in the condensed phase.  
We
consider here the isothermal system without trapping of neutrinos for
simplicity.  
We treat nucleons as a kind of background for kaons and do not take
into account their dynamical degrees of freedom except the kinetic 
energy, which may follow the
spirit by the heavy-baryon chiral-perturbation theory \cite{jen}. 
Relativistic effects for nucleons are beyond the scope of this Letter.

We start with the Kaplan-Nelson Lagrangian with the Goldstone field
$U$ and the Baryon octet $B$ as an effective chiral
Lagrangian ${\cal L}_{chiral}$ \cite{lee,geo},
which consists of the symmetric part,
\beq
{\cal L}_0=\frac{f^2}{4}{\rm tr}(\partial_\mu U^\dagger\partial^\mu U)
+i{\rm tr}\{\bar B (\not D-m_B) B\}
+D{\rm tr}(\bar B\gamma_\mu\gamma_5\{A^\mu, B\})
+F{\rm tr}(\bar B\gamma_\mu\gamma_5[A^\mu, B]).
\label{bb}
\eeq
and the breaking part,
\begin{eqnarray}
{\cal L}_{SB}&=&\alpha{\rm tr}\hat m_q(U+U^\dagger-2) \nonumber\\
&+& a_1{\rm tr}\bar B(\xi \hat m_q\xi+ h.c.)B
+a_2{\rm tr}\bar BB(\xi\hat m_q\xi+h.c.)
+a_3\{{\rm tr}\bar BB\}{{\rm tr}(\hat m_q U+h.c.)}
\label{bc}
\end{eqnarray}
with the quark mass-matrix, $\hat m_q\simeq{\rm diag}(0,0,m_s)$. The
coefficients $\alpha, a_i$ measure the strength of the explicitly
symmetry breaking: the kaon mass is given as $m_K^2\simeq 2m_s\alpha/f^2$
and the $KN$ sigma terms as  
$\Sigma_{Kp}=-m_s(a_1+a_2+2a_3)$ and $\Sigma_{Kn}=-m_s(a_2+2a_3)$. 
The transformation properties of the fields $U$ and $B$
under  $SU(3)_L\times SU(3)_R $ are found in ref.\cite{geo}.

It is well-known that chemical potentials, which
is needed to
ensure various conservation laws in the ground state,  can be
introduced as the artificial ``gauge fields''.
Here we must take into account the conservation laws of two quantities:
electromagnetic charge and baryon number. Accordingly we replace the
time-derivatives by the ``covariant derivatives'';
$
{\cal D}_{\mu=0} U=\partial U/\partial t+i\mu_K[T_{em},
U],
{\cal D}_{\mu=0} B=\partial B/\partial t-i\mu_n B+i\mu_K [T_{em}, B],
$
with the chemical potentials $\mu_K, \mu_n$, and 
the charge operator, $T_{em}\equiv T_3+1/\sqrt{3}T_8
={\rm diag}(2/3, -1/3,-1/3)$ ($T_a$: generators of $SU(3)$). 
Thus the partition function in the imaginary-time formalism can
be represented as follows ($\tau=it, \beta=1/T$);
\beq
  Z_{chiral}=N\int [dU][dB][d\bar B] 
\exp\left[\int_0^\beta d\tau\int d^3x({\cal L}_{chiral}+\delta
{\cal L})\right].
\label{pat}
\eeq
Here we can see that there appears the additional term in the
effective action, $\delta {\cal L}$, 
which is
in general {\it non-invariant} under the chiral transformation, 
\begin{eqnarray}
\delta {\cal L}&=& -\frac{f^2\mu_K}{4}{\rm tr}\{[T_{em}, U]
\dot U^\dagger+
\dot U[T_{em}, U^\dagger]\}
-\frac{\mu_K}{2}{\rm tr}
\{B^\dagger[(\xi^\dagger[T_{em}, \xi]+\xi[T_{em}, \xi^\dagger]), B]\}
\nonumber\\
&-&\frac{f^2\mu_K^2}{4}{\rm tr}\{[T_{em}, U][T_{em}, U^\dagger]\}
+\mu_n{\rm tr}\{B^\dagger B\}-\mu_K{\rm tr}\{B^\dagger[T_{em},B]\},
\end{eqnarray}
where $\dot U\equiv \partial U/\partial\tau$. 

Fluctuations around the condensate are introduced as follows.
First consider the chiral transformations on the 
$SU(3)_L\times SU(3)_R/SU(3)_V\simeq SU(3)$ manifold. Then 
the eight Goldstone fields $\phi_a$ span the coordinates on the
manifold \cite{wei}, $U(\phi_a)=\exp(2iT_a\phi_a/f)\in SU(3)$, 
and accordingly the kaon condensed phase 
corresponds to the point on the manifold specified by 
$U_K=\exp[2i\{T_4\langle\phi_4\rangle
+T_5\langle\phi_5\rangle\}/f]$. 
\footnote{The symbol $\langle {\cal O}\rangle$ implies the thermal
average of an operator ${\cal O}$.}
On the other hand we know that it 
is attained by
the chiral transformation from the vacuum, $U_V=U(\phi_a=0)=1$,
\beq
U_K(\langle\theta\rangle)=\zeta U_V\zeta=\zeta^2
\label{uk}
\eeq 
where $\zeta=\exp(i\langle M\rangle/\sqrt{2}f)$
with the kaon matrix
\[
M=\left[
\begin{array}{ccc}
0 & 0 & K^+ \\
0 & 0 & 0\\
K^- & 0 & 0
\end{array}
\right]
,\qquad K^{\pm}=(\phi_4\pm i\phi_5)/\sqrt{2}\qquad {\rm and}\qquad
\theta^2\equiv 2K^+K^-/f^2. \hskip 3cm
\]

Thorsson and Ellis 
introduced the kaon fluctuation
fields $\tilde K^\pm $ in the usual manner \cite{tho},
$
K^\pm=\langle K^\pm\rangle+\tilde K^\pm. 
$
Here we introduce the fluctuation fields by extending the idea of
chiral rotation mentioned above. First, we generate 
fluctuations $\phi_a$ around the vacuum by a chiral transformation $U_f$,
$
U_f=\eta U_V \eta=\eta^2
$
with $\eta=\exp(iT_a\phi_a/f)$. Then an additional chiral rotation by
$\zeta$ transforms $U_f$ into the form,
\beq
U(\langle\theta\rangle, \phi_a)=\zeta U_f(\phi_a)\zeta.
\label{m}
\eeq
This is our ansatz for the form of $U$ on the manifold.
It is obvious that in the limit $\phi_a\rightarrow 0$ or 
$\langle\theta\rangle\rightarrow 0$, the matrix $U$
is reduced to the relevant form.
We can regard this procedure as a separation 
of the ``zero-mode'' from the full $SU(3)$ matrix $U$ \cite{gass,ty}. 
Accordingly,
the $\xi$ operator, which is defined by $U=\xi^2$, can be obtained
by solving the subsidiary equation,
\beq
\xi=\zeta \eta u^\dagger=u \eta \zeta,
\label{u}
\eeq 
where the matrix $u$ is defined by the second equality in Eq.(\ref{u}).

Defining a new baryon field $B'$ by the use of  the matrix $u$,
$ B'=u^\dagger B u$,
we can see that the chiral-invariant pieces of the Lagrangian are
not changed in form and only the symmetry breaking pieces 
play crucial roles,   
\begin{eqnarray}
{\cal L}(U, B)={\cal L}_0(U, B)+{\cal L}_{SB}(U, B)&\longrightarrow&
{\cal L}_0(U_f, B')+{\cal L}_{SB}(\zeta U_f\zeta,u B'
u^\dagger)\nonumber\\
\delta{\cal L}(U, B)&\longrightarrow&\delta{\cal L}(\zeta U_f\zeta,u B'
u^\dagger).
\label{sb}
\end{eqnarray}
Thus all the dynamics among the condensate, fluctuations and baryons 
are completely prescribed by the non-invariant terms under the
chiral transformation, ${\cal L}_{SB}$ and $\delta {\cal L}$.

To proceed we have to find the $SU(3)$ matrix $u$ by Eq.(\ref{u}). 
It depends on $\phi_a, 
\langle\theta\rangle$ in a complicated nonlinear way, and thereby it is 
very difficult to
find $u$ for the general form of $U_f$ \cite{geo}. 
However, if we 
restrict ourselves to the fluctuations of kaon sector s.t.
\beq
U_f=\exp(i\sqrt{2}M/f),
\eeq
then we can simply find 
\beq
u={\rm diag}(\kappa^*/|\kappa|, 1, \kappa/|\kappa| ),
\label{uu}
\eeq
with 
\beq
\kappa=\cos(\langle\theta\rangle/2)\cos(\theta/2)
-\frac{K^+\langle K^-\rangle}{|\langle K^-\rangle||K^+|}
\sin(\langle\theta\rangle/2)\sin(\theta/2).
\eeq

For the integration measure in Eq.(\ref{pat}), it is {\it invariant} under the
transformation (\ref{m}) by $\zeta$,
$
[dU]\rightarrow [dU_f],
$
and further approximated by the one with flat curvature \cite{gass}.
Thus we find
\beq
Z_{chiral}\simeq \int\prod_{i=1}^8[d\phi_i]
[dB'][d\bar B']\exp[S_{chiral}^{eff}(\zeta, U_f, B')].
\label{z}
\eeq

In the following calculation we only consider the kaon dynamics in
nuclear matter ($B\rightarrow \psi$) in the one-loop order. As mentioned before, since we discard 
the dynamical degrees of freedom of
nucleons in this Letter, we can integrate out 
the kaon field and the nucleon field
separately in Eq. (\ref{z}). First we consider the integration 
over the kaon field.
Using Eq. (\ref{uu}) and retaining only the quadratic terms with respect to the
kaon field in Eq. (\ref{sb}) , we can write the effective action as
\beq
S^{eff}_{chiral}=S_c+S_K^{eff}
\eeq
apart from the nucleon terms, with the tree-level contribution, 
\beq
S_c=\beta
V\left[\frac{1}{2}\mu_K^2f^2\sin^2\langle\theta\rangle
-f^2(m_K^2-\sigma-2\mu_Kb)(1-\cos\langle\theta\rangle)
\right]
\label{cla}
\eeq
and the fluctuation part,
\begin{eqnarray}
S_K^{eff}=\int_0^\beta d\tau\int &d^3x&\Bigl [-\left(\dot K^-
-\mu_K\cos\langle\theta\rangle  K^-\right)
\left(\dot K^+
+\mu_K\cos\langle\theta\rangle K^+\right)
-\mbox{\boldmath$\nabla$} K^+ \mbox{\boldmath$\nabla$} K^- \nonumber\\
&&-\cos\langle\theta\rangle(m_K^2-\sigma) K^+ K^- 
-\frac{\mu_K^2}{4}\sin^2\langle\theta\rangle(K^++K^-)^2 \nonumber\\
&-&b
\left\{K^+(\dot K^--\mu_K\cos\langle\theta\rangle K^-)
-K^-(\dot K^++\mu_K\cos\langle\theta\rangle K^+)\right\}+O(|K|^4)
\Bigr ],
\end{eqnarray}
where $b$ stands for the V-spin density, $b=
\langle\bar\psi\gamma_0\frac{1}{8}(3+\tau_3)\psi\rangle/f^2$
and $\sigma$ the $KN$ sigma-term contribution, 
$\sigma=\langle\bar\psi\Sigma_{KN}\psi\rangle/f^2$ with 
$\Sigma_{KN}=\left(\begin{array}{cc} \Sigma_{Kp} & 0 \\0 & \Sigma_{Kn}
\end{array}\right)$. 
\footnote{It would
be interesting to compare our result with the one given in
ref.~\cite{tho}; although our expression might look rather simple,
the results, e.g. the dispersion relations for the kaonic modes, are
consistent with them.}  
Note that Eqs.~(14), (15) are the
exact consequences from the general argument given in Eq.~(8). 
Expanding the fluctuation field 
with the Matsubara frequency $\omega_n=2\pi n T$ and performing the
integration, we find the kaon-loop contribution,
\beq
\ln Z_K^{eff}=\sum_{n, {\bf p}}\ln(|D^{eff}|^{-1/2}),
\eeq
with the inverse thermal Green function $D^{eff}$, 
\beq
D^{eff}=\beta^2\left(
\begin{array}{cc}
\omega_n^2+(\tilde\omega_--\tilde\mu_K)
(\tilde\omega_++\tilde\mu_K)+\mu_K^2\sin^2\langle\theta\rangle & 
2(\tilde\mu_K+b)\omega_n\\
-2(\tilde\mu_K+b)\omega_n & 
\omega_n^2+(\tilde\omega_--\tilde\mu_K )
(\tilde\omega_++\tilde\mu_K)
\end{array}
\right),
\label{deff}
\eeq
where 
$
\tilde\omega_\pm=\pm b+(p^2+\tilde m_K^{*2}+b^2)^{1/2}, 
$
$\tilde\mu_K=\mu_K\cos\langle\theta\rangle$ and the effective mass, 
$\tilde m_K^{*2}=\cos\langle\theta\rangle[m_K^2-\sigma]$.
\footnote{In the limit $\langle\theta\rangle\rightarrow 0, 
\tilde\omega_\pm\rightarrow\omega_\pm=\pm b+(p^2+m_K^{*2}+b^2)^{1/2}$,
which is nothing but the dispersion relation for kaons in the normal
phase
\cite{lee,tat}.}

The excitation energy of the kaonic modes is given by the
poles of the thermal Green function.
Then we have two solutions corresponding to the $K^\pm$ modes,
\beq
E_\pm^2=(c_1+c_3+2c_2^2)\pm\sqrt{(c_3+2c_2^2)^2+4c_1c_2^2},
\label{disp}
\eeq
where $c_1=(\tilde\omega_--\tilde\mu_K)
(\tilde\omega_++\tilde\mu_K)$, 
$c_2=\tilde\mu_K+b$  
and $c_3=1/2\mu_K^2\sin^2\langle\theta\rangle$.
In the condensed phase ($\langle\theta\rangle\neq 0$), 
the mode corresponding to $E_-$ is the Goldstone mode as a consequence 
of the breakdown of V-spin symmerty in the ground state; the ground
state is not an eigenstate of the conserved operator $V_3$.
It is easy to show $E_-({\bf p}=0)\sim 0$. In the tree-level
approximation, $c_1=0$ as a result of the extremum condition for $S_c$, 
$\partial S_c/\partial \langle\theta\rangle=0$, which means $E_-=0$, 
while we should find 
$E_-({\bf p}=0)=0+O(\hbar)$ once taking into account
the fluctuation effects (kaon loops) \cite{tho}. 
When we consider the thermal kaon loops, $E_-$ directly enters into the
Bose-Einstein distribution function and it should diverge at ${\bf
p}=0$. The other is the massive mode and can be neglected for
temperature we are interested in $(T\leq 100$MeV).

Moreover, if we can safely neglect the $c_3$ term, we get a simple expression
for $E_\pm$,
\beq
E_\pm=\tilde\omega_\pm\pm\tilde\mu_K
=\sqrt{p^2+\tilde m_K^{*2}+b^2}\pm(b+\tilde\mu_K).
\label{dispa}
\eeq
We can
expect the pertinence of this approximated formula qualitatively by
several reasons: first, this term includes the $KK$ scattering
amplitude in the
leading order in difference from the other terms in Eq. (\ref{deff}), 
which should be small at low-energy regime due to the
Goldstone-particle nature of kaons. 
Secondly, in the limit, $\langle\theta\rangle\rightarrow
0$ or $\mu_K\rightarrow 0$, $c_3$ gives no contribution. We also know
that $\langle\theta\rangle$ and $\mu_K$ are inversely proportional to each
other referring the previous studies, e.g., \cite{lee,tat,thor}. 
Leaving the detailed discussions in ref.~\cite{ty}, we find that the
approximation is justified if the following condition is satisfied,
\beq
c_3\ll \frac{C\pi}{16}T\zeta^2(3/2)\quad 
{\rm or} \quad 
\mu_K\sin\langle\theta\rangle\ll \frac{1}{4}(2\pi CT)^{1/2}\zeta(3/2),
\label{con}
\eeq
with $C^2=\tilde m_K^{*2}+b^2$.
We shall see, by
numerical calculations, that this
approximation really works well even at finite temperature except very 
low temperature, where thermal effects should be trivially unimportant.
Finally, it is to be noted that this approximation
never violate the Goldstone-particle nature, $E_-({\bf p}=0)=0+O(\hbar)$. 

Then the thermodynamic potential by the 
kaon-loop contribution, $\Omega_K^{eff}=-T\ln Z_K^{eff}$,  reads
\beq
\Omega_K^{eff}=V\int \frac{d^3 p}{(2\pi)^3}\frac{1}{2}
[E_+({\bf p})+E_-({\bf p})]+TV\int \frac{d^3 p}{(2\pi)^3}
\ln(1-e^{-\beta E_+({\bf p})})(1-e^{-\beta E_-({\bf p})}).
\eeq
The first term is the zero-point energy (ZP) contribution and 
the second one the thermal one. Since ZP contribution includes 
the divergent integrals, they should be
renormalized in some way. Following the method proposed 
in ref.~\cite{tho}, we can extract
the finite contribution $\Omega^{ZP,r}$; 
under the approximation, $c_3=0$, $\Omega^{ZP,r}$ takes a simple
form,
\beq
\Omega^{ZP,r}\sim-\frac{1}{64\pi^2}V
\left[(m_K^2-3C^2)(m_K^2-C^2)+2C^4\ln\frac{m_K^2}{C^2}\right].
\label{zpr}
\eeq
We shall see numerically that $\Omega^{ZP,r}$ results in a tiny value, 
as consistent with  ref.~\cite{tho}, we discard it in the following.

In the rest of this Letter we present some results about the
dispersion relation and EOS at finite temperature.
For nucleons we should take into account at least two effects, otherwise the
results should become meaningless. First one is the thermal effects for the
kinetic energy term; it is given by the standard formula,
\beq
\Omega_N^{kin}\simeq-2TV\sum_{n,p}\int\frac{d^3
p}{(2\pi)^3}\ln(1+e^{-\beta(\epsilon-\mu_i)}),
\eeq
where $\epsilon=p^2/2M$ for nonrelativistic nucleons and $\mu_p$ is 
defined by the relation, $\mu_p=\mu_n-\mu_K$. Secondly, it is well-known that 
the nuclear symmetry energy plays an important role for the
ground-state properties of the condensed phase. Since we have already included 
the kinetic energy for nucleons, we take into account only the potential
energy contribution. Following Prakash et al. \cite{prak} we
effectively introduce a symmetry energy contribution,
\beq
\Omega_N^{symm}=V\rho_B(1-2x)^2S^{pot}(u),
\eeq  
where $\rho_B$ the baryon number density, 
$u=\rho_B/\rho_0$ ($\rho_0\simeq 0.16$fm$^{-3}$ the nuclear
density), and $x$ is the proton-number fraction. 
The coefficient $S^{pot}(u)$ is given as
$
S^{pot}(u)=(S_0-(2^{2/3}-1)(3/5)E_F^0)F(u)
$
with the empirical symmetry energy $S_0\simeq 30$MeV 
and the Fermi energy $E_F^0$ at $\rho_0$. 
We take here $F(u)=u$ as a form of the control function for simplicity.
Thus the nucleon contribution is given by 
$
\Omega_N^{eff}=\Omega_N^{kin}+\Omega_N^{symm}.
$

Finally, the total thermodynamic potential  $\Omega_{total}$ 
is given by further adding the one for
leptons (electrons and muons),
$\Omega_l$,
\beq
\Omega_{total}=\Omega_c+\Omega_K^{eff}+\Omega_N^{eff}+\Omega_l,
\label{omega}
\eeq
where $\Omega_c$ is the tree-level contribution, $\Omega_c=-TS_c$,
given by Eq. (\ref{cla}) and we can
use the formulae of free leptons for $\Omega_l$. 
The parameters $x,\langle\theta\rangle$ and the chemical potential 
$\mu_K$
are determined by the extremum conditions for given density and
temperature like in the $T=0$ case, e.g., \cite{lee,tat,thor}:
$\partial\Omega_{total}/\partial x=0, 
\partial\Omega_{total}/\partial\langle\theta\rangle=0$ and
$\partial\Omega_{total}/\partial\mu_K=0$ by the use of Eq.(\ref{omega}).

In the following we present some numerical examples to see how our
formalism works and how large the thermal effects are, leaving the full
discussion about the thermal effects and the parameter dependence 
in a separated paper \cite{ty}. 
In numerical calculations we use the values, $a_1m_s=-67$ MeV, 
$a_2m_s=134$ MeV and $a_3m_s=-222$ MeV \cite{thor}. 
First of all we examine the pertinence of the approximation $c_3=0$. 
We compared the exact
dispersion relation (\ref{disp}) with the approximated one (\ref{dispa}) 
as a
function of momentum up to 500 MeV and checked that this is a fairly good 
approximation. 
\footnote{We also compared our dispersion relations with
ref.\cite{tho} at $T=0$ and found that they are consistent with each
other regardless of the magnitude of $\langle\theta\rangle$ in the
ground state.}
In Fig.1 we show a typical example, which might be the
{\it worst} case at $T=50$ MeV in such a sense that the difference
between them becomes largest. We can see that the difference
is very small; in fact, it causes the error of only less than several 
MeV$\cdot$fm$^{-3}$ for
pressure. As a general trend the difference becomes
smaller as density or temperature becomes higher (c.f. Eq.~(\ref{con}) ).
It should be
worth noting that once this approximation works, the expressions of
thermodynamic quantities like charge, entropy, energy become very
clear \cite{ty}.

In Fig.2 we present an example of EOS for the hot neutron-star
matter by using $c_3=0$. We construct EOS for the condensed phase, 
referring to the one for normal neutron-star matter; $P=P_V+P_{chiral}$.
The pressure $P_{chiral}$ is simply given by Eq.(\ref{omega}) through 
the thermodynamic relation, $P_{chiral}V=-\Omega_{tot}$.
As a reference EOS for normal neutron-star matter we
employ here the one suggested by Prakash et al.\cite{prak}; 
$P_V=u^2\partial(V(u)/u)/\partial u$ with the potential contribution $V(u)$.
The results indicate the first-order phase transition in Fig. 2. Hence,
the Maxwell construction should be imposed to get the real EOS.
We
can easily see that the thermal effects are remarkably larger than those 
in the normal phase; this is caused by the reduction of the chiral
angle, and the soft mode mainly contributes to this effect. 

Finally we have checked the smallness of the ZP contribution to be
tiny; the
value of the ZP contribution to pressure, $P_{ZP}=-\Omega^{ZP, r}/V$,
is always less than several MeV$\cdot$fm$^{-3}$ 
over the density-temperature region we are interested in.

In this Letter we have presented a formalism to treat fluctuations around 
the condensate within the framework of chiral symmetry. 
Our approach is based on the group theoretical
argument; Goldstone fields are regarded as the coordinates on the
chiral $SU(3)_L\times SU(3)_R/SU(3)_V$ manifold and the kaon condensed
phase can be
represented as a chiral rotated state from the vacuum. Using this idea
we have introduced 
fluctuations around the condensate by way of the
successive chiral transformations.
We derived an effective action and extracted the dispersion relation 
for kaonic excitations. 
There appear two modes:
Goldstone-like soft mode and very massive one, which
correspond to $K^-$ and $K^+$ mesonic excitations,
respectively. 
We have discussed that the form of
the dispersion relation can be reduced to a simple one by 
the approximation, $c_3=0$, which  
was also checked numerically and proved to be fairly good.

Applying our formalism to the derivation of EOS as an example, 
we have pointed out the largeness of the thermal effects 
in the condensed phase; they 
result in a remarkable reduction of pressure in the condensed phase,
compared with the normal phase. 

In this Letter we have not taken into account the dynamical degrees of 
freedom of nucleons consistently, but it is not difficult to treat them
consistently based on the idea given in this Letter \cite{ty}.

\vskip 1cm

We thank T. Muto for discussions and careful reading of this
manuscript.
This work was supported in part by the Japanese Grant-in-Aid for
Scientific Research Fund of the Ministry of Education, Science and
Culture (08640369).

\newpage
\centerline{{\Large Figure Captions}}

\bigskip

\begin{itemize}

\item[Fig. 1]
Dispersion relation for kaonic modes in the condensed phase at
$u=3.87$ and $T=50$ MeV. Solid lines show the results given by 
Eq.~(\ref{dispa}), while dotted lines by Eq.~(\ref{disp}).
\item[Fig. 2]
Equation of state for neutron-star matter at finite temperature. EOS
at T=0 , which is the same one given in ref. \cite{thor},  is also
shown for comparison. Dotted lines show EOS for normal
neutron-star matter. 
\end{itemize}


\clearpage
\begin{thebibliography}{MMMM}

\bibitem{lee} D.B. Kaplan and A.E. Nelson, Phys. Lett. {\bf B175}
(1986) 57 ; {\bf B179} (1986) 409(E).\\
For a recent review, 
C.-H. Lee, Phys. Reports {\bf 275} (1996) 197.

\bibitem{bb} G.E. Brown and H.A. Bethe, Astrophys. J. {\bf 423}
(1994) 659.

\bibitem{bau} T.W. Baumgarte, S.L. Shapiro and S. Teukolsky,
Astrophys. J. {\bf 443} (1995) 717; {\bf 458} (1996) 680.

\bibitem{tat} H. Fujii, T. Maruyama, T. Muto and T. Tatsumi, 
Nucl. Phys. {\bf A597}, 645 (1996).\\
As a review, T. Tatsumi, Prog. Theor. Phys. Suppl. {\bf 120} (1995) 111    
and references cited therein.

\bibitem{thor} V. Thorsson, M. Prakash and J.M. Lattimer,
Nucl. Phys. {\bf A572} (1994) 693; {\bf A574} (1994) 851.
 
\bibitem{kei} W. Keil and H.-Th. Janka, Astron. Astrophys. {\bf 296}
(1995) 145.

\bibitem{pra} M. Prakash, I. Bombaci, M. Prakash, P.J.Ellis,
J.M. Lattimer, R. Knorren, Phys. Reports {\bf 280} (1997) 1.

\bibitem{tho} V. Thorsson and P.J. Ellis, Phys. Rev. {\bf D55} (1997)
5177.

\bibitem{jen} E. Jenkins and A. Manohar, Nucl. Phys. {\bf B368} (1992) 
190.

\bibitem{geo} H. Georgi, {\it Weak interactions and modern particle
theory} (Benjamin/Cummings Pub., 1984).

\bibitem{wei} S. Weinberg, Phys. Rev. {\bf 166} (1968) 1568; Physica
{\bf 96A}
(1979) 327; {\it The quantum theory of field II} (Cambridge U. Press, 1996).

\bibitem{gass}J. Gasser and H. Leutwyler, Phys. Lett. {\bf B188}
(1987) 477.\\
F.C. Hansen, Nucl. Phys. {\bf B345} (1990) 685.

\bibitem{prak} M. Prakash, T.L. Ainsworth and J.M. Lattimer, Phys
Rev. Lett. {\bf 61} (1988) 2518.

\bibitem{ty} T. Tatsumi and M. Yasuhira, in preparation.
\end{thebibliography}
\end{document}